\documentclass[a4paper,11pt]{article}
    \usepackage{jheppub} 
    \usepackage{lineno}

    \title{\boldmath Measurements of inclusive $J/\psi$ production in Au+Au collisions at $\sqrt{s_{\rm NN}}$ = 14.6, 17.3, 19.6, and 27 GeV with the STAR experiment}
    
    \collaboration{The STAR Collaboration}
    
    \abstract{Inclusive $J/\psi$ production at mid-rapidity ($|y| < 1.0$) in Au+Au collisions at center-of-mass energies per nucleon-nucleon pair ($\sqrt{s_{\rm NN}}$) of 14.6, 17.3, 19.6, and 27 GeV is measured with the STAR detector at the Relativistic Heavy Ion Collider. The nuclear modification factor $R_{{\rm AA}}$ is used to quantify modifications of $J/\psi$ production relative to an interpolated $p$+$p$ reference. The measured $R_{{\rm AA}}$ shows a hint of a decreasing trend from peripheral to central collisions, reaching a value of approximately 0.5 in the 0--20\% centrality class across all collision energies. Unlike similar measurements in Au+Au collisions at $\sqrt{s_{\rm NN}}$ = 200 GeV, where $J/\psi$ $R_{{\rm AA}}$ shows little dependence on transverse momentum ($p_{\rm {T}}$), the corresponding $R_{{\rm AA}}$ at the lower collision energies of 14.6--27 GeV shows a tendency to increase with $p_{\rm {T}}$. Additionally, the inclusive $J/\psi$ $R_{{\rm AA}}$, integrated over $J/\psi$ $p_{\rm {T}}$, in central collisions exhibits no significant variation within uncertainties over the energy range of 14.6 to 200 GeV. These results can be qualitatively described by transport models that incorporate hot and cold nuclear matter effects.}
    
\begin{document}
    \maketitle
    \flushbottom
    
    \section{Introduction}
    \label{introduction}       
    
    Quantum Chromodynamics (QCD) predicts that at extremely high temperatures and nuclear densities, nucleons may transition into a state of deconfined quarks and gluons, forming a Quark-Gluon Plasma (QGP)~\cite{Harris:1996zx,Busza:2018rrf,Harris:2023tti}. This deconfined state existed in the early universe shortly after the Big Bang and can be recreated in the laboratory with high-energy heavy-ion collisions. Understanding the properties of the QGP is essential for exploring the fundamental characteristics of the strong interaction, making it one of the primary research objectives at facilities such as the Relativistic Heavy Ion Collider (RHIC) and the Large Hadron Collider (LHC)~\cite{STAR:2005gfr,PHENIX:2004vcz,PHOBOS:2004zne,BRAHMS:2004adc,ALICE:2022wpn,CMS:2024krd,ATLAS:2018hqe,Chen:2026gka,Chen:2024aom}.
    
    Charmonia are bound states of charm-anticharm ($c\bar{c}$) pairs and serve as valuable probes of the QGP created in relativistic heavy-ion collisions. Due to their large masses, $c\bar{c}$ pairs are primarily produced in the initial partonic scatterings before the QGP formation, and can subsequently form charmonium states. Charmonia can dissociate in the QGP through two possible mechanisms: screening of the binding potential by color charges~\cite{Matsui:1986dk}, and scatterings with medium constituents~\cite{Andronic:2025jbp}. The dissociation effect results in a reduction in charmonium production compared to the case without QGP effects. On the other hand, deconfined charm and anticharm quarks can regenerate charmonia, compensating for the yield loss due to dissociation. Together, the dissociation and regeneration processes reflect the medium's properties. In addition, cold nuclear matter (CNM) effects, such as modifications of parton distributions in nuclei compared to those in free nucleons (nPDF effects), initial-state energy loss, and nuclear absorption, also alter charmonium production in heavy-ion collisions. For example, nPDF effects include both shadowing and antishadowing contributions, while initial-state energy loss and nuclear absorption tend to reduce the yield~\cite{Nagle:2010ix, Nagle:1999ms, Antreasyan:1978cw, Vogt:2001ky, Ferreiro:2014bia}.
    
    Among the various charmonium states, the $J/\psi$ meson is the easiest one to access experimentally and has been measured extensively, including inclusive $J/\psi$ production in Au+Au collisions at center-of-mass energies per nucleon-nucleon pair ($\sqrt{s_{\rm NN}}$) of 39, 54.4, 62.4, and 200 GeV~\cite{STAR:2016utm,STAR:2025lxx,STAR:2019fge} at RHIC, in Pb+Pb collisions at $\sqrt{s_{\rm NN}} = 17.3$ GeV~\cite{NA50:2000brc} at the Super Proton Synchrotron (SPS), and at 2.76 TeV and 5.02 TeV at the LHC~\cite{Hushnud:2023gyl,ALICE:2013osk,Bai:2020svs}. Here, inclusive $J/\psi$ production includes four main sources: directly produced $J/\psi$ from initial partonic scatterings, regenerated $J/\psi$, decays of excited charmonium states such as $\chi_c$ and $\psi(2S)$, and decays of long-lived $B$ hadrons (denoted ``non-prompt''). Together, these measurements exhibit a non-trivial dependence on collision energy, reflecting the changing relative importance of dissociation, regeneration, and CNM effects. In the LHC energy regime, the large charm quark production cross section leads to a significant regeneration contribution to $J/\psi$ production, particularly at low $p_{\rm T}$~\cite{ALICE:2013osk,Bai:2020svs}.  Toward lower collision energies, the charm quark production cross section and medium temperature decrease, leading to weaker regeneration and dissociation effects~\cite{Andronic:2025jbp,Zhao:2020jqu,Zhao:2022ggw}. On the other hand, nuclear absorption and other CNM effects are expected to become relatively more important with decreasing energy~\cite{Lourenco:2008sk,Andronic:2025jbp}. However, existing data leave an unexplored energy range between 17.3 GeV and 39 GeV, which is covered by the Beam Energy Scan Phase II (BES-II) program at RHIC. This provides a useful opportunity to study the transition from a regime where regeneration can be significant to one where CNM effects play a larger role. 
    
    In this article, new measurements of inclusive $J/\psi$ production as functions of transverse momentum ($p_{\rm {T}}$) and collision centrality in Au+Au collisions at $\sqrt{s_{\rm NN}} =$ 14.6, 17.3, 19.6, and 27 GeV by the Solenoidal Tracker at RHIC (STAR) experiment are presented. These measurements fill the gap between 17.3 and 39 GeV in currently available data and extend the energy reach down to 14.6 GeV. They provide new constraints on the energy evolution of $J/\psi$ production and aid in the disentanglement of dissociation, regeneration and CNM effects.
    
    \section{Experiment and analysis}
    The datasets used in this analysis consist of Au+Au collisions at $\sqrt{s_{\rm NN}} = 14.6$, 17.3, 19.6, and 27 GeV, collected in 2018, 2019, and 2021 by the STAR experiment~\cite{STAR:2002eio} at RHIC. The primary detector subsystems employed include the Time Projection Chamber (TPC)~\cite{Anderson:2003ur, Shen:2018pkc} and Time of Flight (TOF) detector~\cite{Llope:2012zz}, covering mid-pseudorapidity ($|\eta|$ $<$ 1), and trigger detectors at forward rapidities.
    
    A minimum-bias trigger is used for data taking. For the 14.6, 17.3, and 19.6 GeV data, the trigger requires an east-west coincidence in at least one of the Vertex Position Detector (VPD)~\cite{Llope:2014nva}, Zero Degree Calorimeter (ZDC)~\cite{Adler:2000bd}, or Event Plane Detector (EPD)~\cite{Adams:2019fpo} detector systems. For the 27 GeV data, the corresponding trigger uses the VPD, ZDC, or Beam-Beam Counter (BBC)~\cite{Bieser:2002ah}. To ensure good TPC acceptance, the collision vertex, reconstructed using tracks found in the TPC, is required to lie within 70 cm of the nominal interaction point along the $z$-axis (beam direction) for 27 GeV collisions and within 145 cm for 14.6, 17.3, and 19.6 GeV collisions. The different $V_z$ requirements follow the event selection criteria and centrality definitions used for the corresponding data sets; the wider requirement at 14.6, 17.3, and 19.6 GeV retains sufficient numbers of events for these measurements. To suppress background from the beam hitting the beam pipe, the radial position of the reconstructed vertex is required to be at most 2 cm from the central axis of STAR. Pile-up events, where multiple collisions occur within the TPC readout time window, are removed based on the correlation between the efficiency-uncorrected charged-particle multiplicity in the TPC and the number of TPC tracks matched to the TOF, since the TOF is a fast detector resilient to pile-up effects. Here, the charged-particle multiplicity is defined as the number of tracks measured in the TPC within $|\eta| < 0.5$ and with a distance of closest approach (DCA) to the collision vertex less than 1 cm. After event selection, a total of about $392$, $282$, $689$, and $509$ million events are retained for 14.6, 17.3, 19.6, and 27 GeV Au+Au collisions, respectively.
    
High-quality tracks are selected by requiring that at least 20 TPC hits were used for track reconstruction at 27 GeV and at least 40 TPC hits at lower energies. The larger number of required TPC hits at lower energies is made possible by the upgrade of the inner part of the TPC~\cite{Shen:2018pkc}, which provides more space points per track and allows for improved tracking and particle identification (PID) performance. To suppress split tracks, the ratio of the number of TPC hits used for track reconstruction to the maximum possible number of hits along the track trajectory is required to be at least 0.52. The track DCA to the collision vertex is required to be at most 1 cm to reduce contamination from secondary decays. Tracks are required to have $p_{\rm T}\geq$ 0.2 GeV/$c$, as those with lower $p_{\rm T}$ cannot be reliably reconstructed in the TPC. In addition, tracks should lie within $|\eta|\leq1$. The specific energy loss ($dE/dx$) of charged particles in the TPC provides essential information for PID. To ensure good $dE/dx$ resolution, at least 15 TPC hits are required for the $dE/dx$ calculation at 27 GeV, and at least 30 are required at 14.6, 17.3, and 19.6 GeV.
    
Collision centrality is used to classify the collision geometry. Central (peripheral) events refer to those with small (large) impact parameters and large (small) nuclear overlaps. Centrality determination is based on the measured charged-particle multiplicity distribution with corrections for dependences on luminosity and collision vertex position. By fitting the multiplicity distribution in data to the Glauber model~\cite{Miller:2007ri}, centrality classes are determined, and the average number of binary nucleon-nucleon collisions ($\langle N_{\rm {coll}} \rangle$) and number of participants ($\langle N_{\rm {part}} \rangle$) for each centrality class are also extracted.
    
    \section{Electron identification and $J/\psi$ reconstruction}
    
    Electron identification (eID) is based on $dE/dx$ measured in the TPC and the arrival time measured by the TOF.

The normalized variable $n\sigma_e$, which quantifies the number of standard deviations by which a track's $dE/dx$ deviates from the expected value for electrons, is used for eID. It is defined as:
\begin{equation}
n\sigma_e = \frac{1}{R_{dE/dx}} \ln \left( \frac{\langle dE/dx \rangle^{\rm Meas.}}{\langle dE/dx \rangle^{\rm Th.}_e} \right),
\label{Eq:nsigmae}
\end{equation}
where $\langle dE/dx \rangle^{\rm Meas.}$ is the measured energy loss, $\langle dE/dx \rangle^{\rm Th.}_e$ is the expected value for electrons based on the Bichsel formalism~\cite{Bichsel:2006cs,Bichsel:1988if}, and $R_{dE/dx}$ is the resolution of $\ln(dE/dx)$. The $n\sigma_e$ distribution for electrons is expected to follow a Gaussian distribution with a mean of zero and a standard deviation of one.

The value of $\beta = v/c$, where $v$ denotes the particle velocity and $c$ is the speed of light, is obtained by combining the arrival time measured by the TOF, the collision start time measured by the VPD, and the path length measured in the TPC. Owing to their small mass, $1/\beta$ of electrons is approximately equal to 1, and a selection of $|1/\beta - 1| < 0.025$ is used to remove hadronic background.
    
\begin{figure}[htbp]
    \centering
    \includegraphics[width=0.6\textwidth]{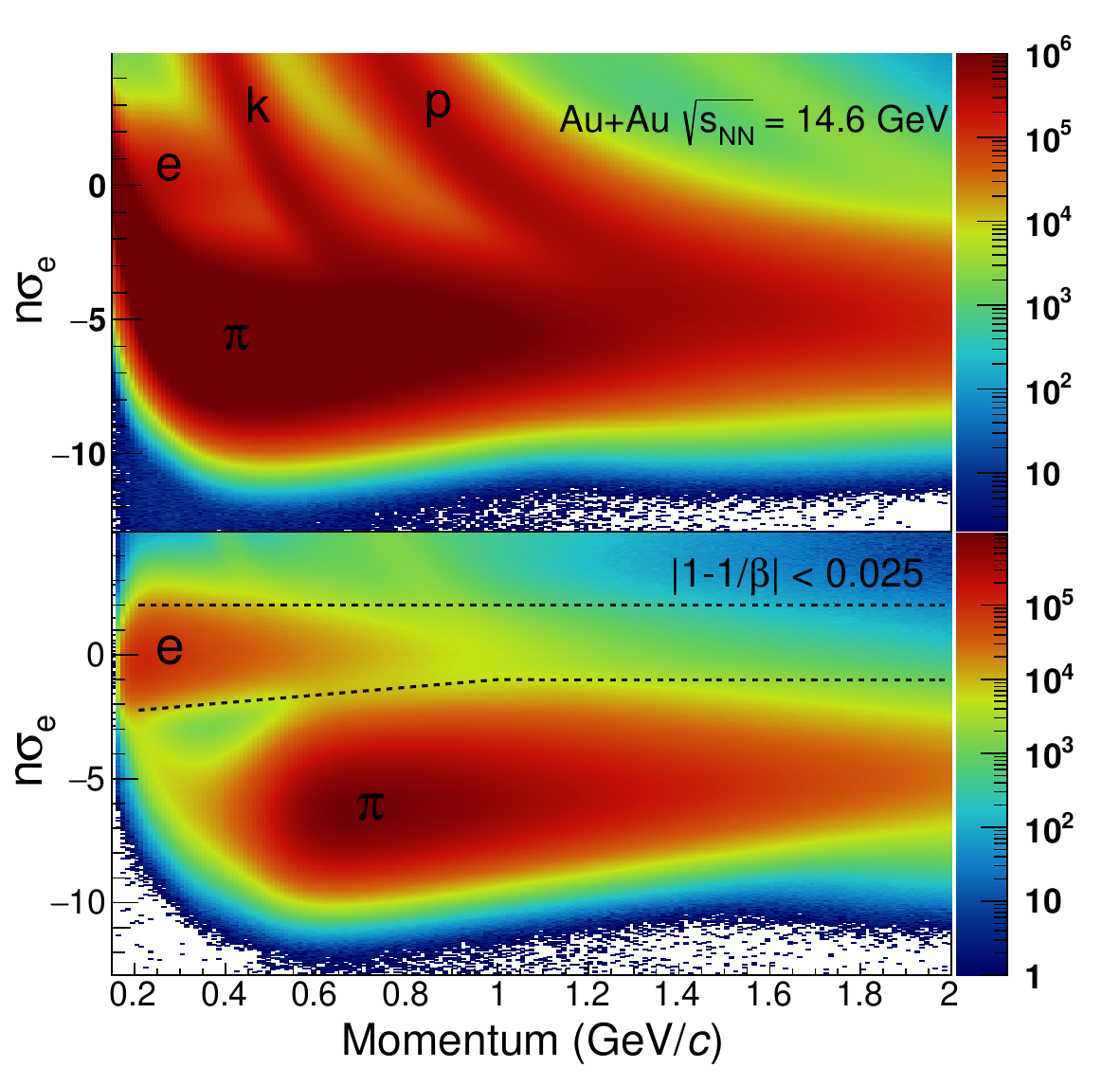}
    \caption{$n\sigma_e$ distribution as a function of charged-track momentum in Au+Au collisions at $\sqrt{s_{\rm NN}} = 14.6$ GeV. The top and bottom panels show the distributions before and after applying the $|1/\beta - 1| < 0.025$ selection. The dashed lines indicate the $n\sigma_e$ selection criteria employed in the analysis.}
    \label{Fig:nsigmae_vs_p}
    \end{figure}

    The top and bottom panels of Fig.~\ref{Fig:nsigmae_vs_p} show an example of the $n\sigma_e$ distribution as a function of momentum for charged-particle tracks before and after the $|1/\beta - 1|  < 0.025$ requirement in Au+Au collisions at $\sqrt{s_{\rm NN}}=14.6$ GeV. After applying the $1/\beta$ selection, a clear electron band is observed, and a momentum-dependent $n\sigma_e$ selection criterion, indicated by the black dashed lines, is used to further remove the hadronic background. Slightly different $n\sigma_e$ selections are used at other energies due to variations in the $dE/dx$ calibration.
    
\begin{figure}
    	\centering 
    	\includegraphics[width=0.9\textwidth]{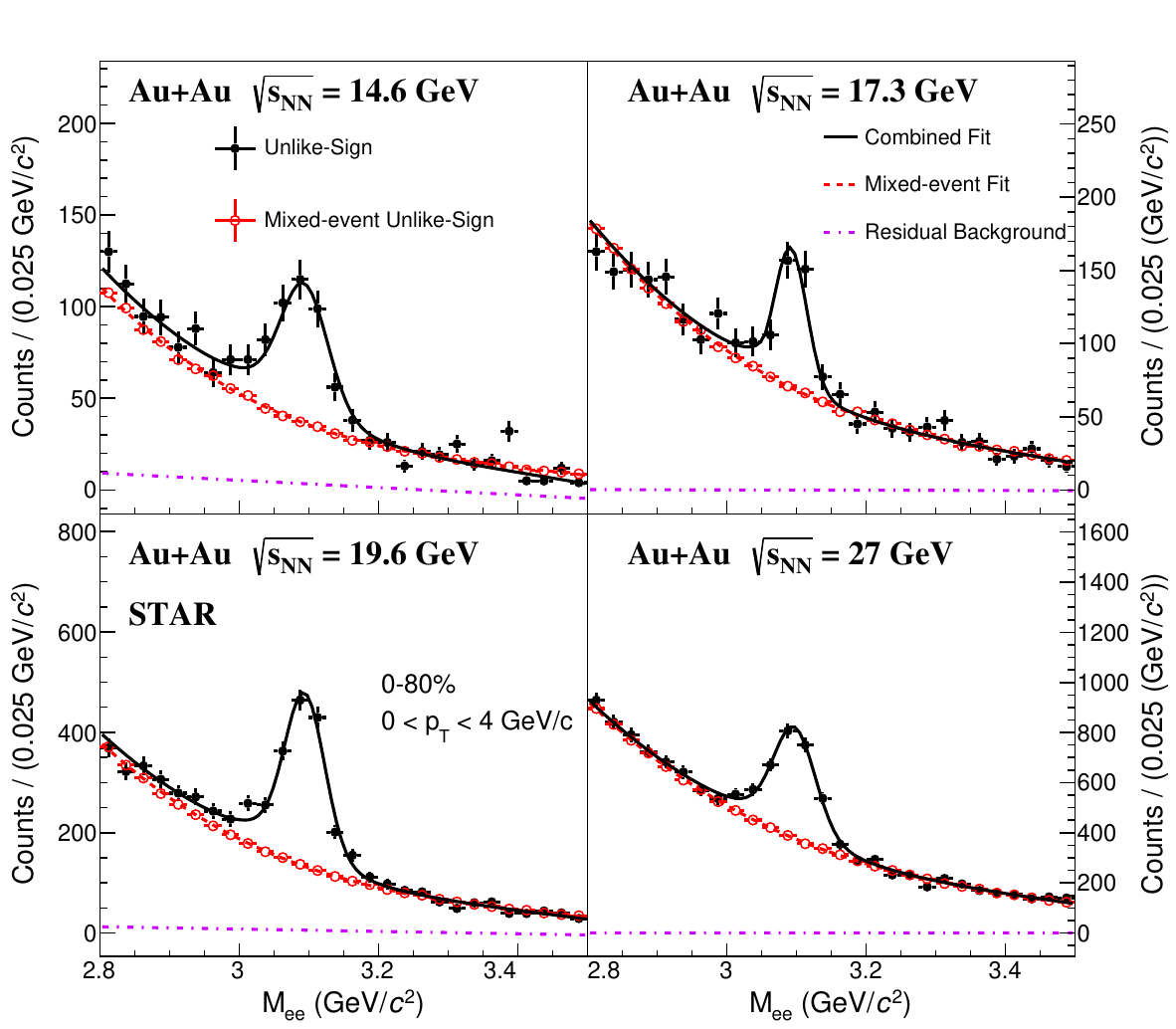}	
        \caption{Distributions of $e^{+}e^{-}$ invariant mass (black solid circles) and the combinatorial background estimated using mixed events (red open circles) in 0--80\% Au+Au collisions at $\sqrt{s_{\rm NN}}$ = 14.6, 17.3, 19.6, and 27 GeV, integrated over $0 < p_{\rm T} < 4$ GeV/$c$ for the $e^{+}e^{-}$ pair. The black solid curve represents the combined fit, which includes the $J/\psi$ signal, the combinatorial background (red dotted curve), and the residual background (purple dash-dotted curve).}
    	\label{Fig:RawSig}%
    \end{figure}

    $J/\psi$ candidates are reconstructed from $e^{+}e^{-}$ pairs. Figure~\ref{Fig:RawSig} illustrates the invariant mass distributions integrated over centrality and $e^{+}e^{-}$ pair $p_{\rm T}$, labeled ``Unlike-Sign'', for the datasets studied.
     Raw $J/\psi$ yields are extracted through a minimum $\chi^2$ fit to the invariant mass spectra. The fit function includes:
    i) the $J/\psi$ signal, modeled by a Crystal Ball function~\cite{Skwarnicki:1986xj} whose shape parameters are derived from simulations tuned to match track momentum resolution in data;
    ii) the combinatorial background, modeled by a third-order polynomial function; and
    iii) the residual background, such as $e^{+}e^{-}$ pairs from heavy-flavor decays and the Drell--Yan process, approximated by a first-order polynomial function. The shape of the combinatorial background is fitted to the $e^{+}e^{-}$ distribution from mixed events, labeled ``Mixed-event Unlike-Sign'' in Fig.~\ref{Fig:RawSig}, which is constructed by pairing $e^{+}$ and $e^{-}$ candidates from different events with similar event properties, including collision centrality and primary vertex position. This mixed-event distribution is normalized by the ratio of the same-event like-sign ($e^+e^+$ and $e^-e^-$) distribution to that in mixed events. The same fitting procedure is applied across different centrality classes and $J/\psi$ $p_{\rm T}$ intervals.
    
    \section{Efficiency and acceptance correction}
    
    
    The $J/\psi$ reconstruction efficiency is evaluated with a toy Monte Carlo simulation that combines the $J/\psi \rightarrow e^{+}e^{-}$ decay kinematics with the reconstruction efficiency for daughter electrons. The single-electron efficiency is defined as
\begin{equation}
\varepsilon_{\rm e} = \varepsilon_{\rm track} \times \varepsilon_{\rm match} \times \varepsilon_{\rm eID},
\end{equation}
where $\varepsilon_{\rm track}$, $\varepsilon_{\rm match}$, and $\varepsilon_{\rm eID}$ are the TPC tracking, TOF matching, and eID efficiencies, respectively. For each generated $J/\psi$, efficiencies of daughter electrons are obtained according to their kinematics and multiplied to form the pair weight. The total $J/\psi$ reconstruction efficiency, $\varepsilon_{\rm total}$, is then obtained in each $J/\psi$ kinematic and centrality interval as the ratio of the efficiency-weighted pair yield to the generated $J/\psi$ yield.
    
    The TPC tracking efficiency ($\varepsilon_{\rm {track}}$), including the TPC response and acceptance, is evaluated using detector simulations. Simulated electrons are propagated through a GEANT3 simulation~\cite{STAR:2008med} of the STAR detector and embedded into real Au+Au events at the raw detector signal level. Here, real events are randomly sampled from the full data set at each energy, and the number of embedded tracks is limited to 5\% of the measured charged-particle multiplicity to avoid distortion of the TPC performance. These embedded events are then processed with the same reconstruction software as used for real data. The TPC tracking efficiency is defined as $\varepsilon_{\rm{track}} = N_{\rm {rec}}/N_{\rm {emb}}$, where $N_{\rm {rec}}$ is the number of reconstructed MC tracks that satisfy track quality requirements, and $N_{\rm {emb}}$ is the number of simulated MC tracks. $\varepsilon_{\rm{track}}$ is evaluated as a function of electron $p_{\rm T}$, $\eta$, and $\phi$.

    The TOF matching efficiency ($\varepsilon_{\rm match}$), which accounts for the TOF response and the acceptance difference between the TPC and TOF, is obtained with a data-driven method. First, a high-purity pion sample is selected using a stringent $dE/dx$ requirement. Using this sample, the TOF matching efficiency for pions is calculated as the ratio of the number of selected pion tracks with a valid TOF hit to that in the TPC. A three-dimensional pion efficiency map as a function of $p_{\rm T}$, $\eta$, and $\phi$ is constructed. Secondly, to account for the species dependence of the TOF matching efficiency, a pure electron sample, also referred to as photonic electrons, is selected from photon conversions and $\pi^0$ Dalitz decays~\cite{STAR:2013pwb} by requiring $e^{+}e^{-}$ pairs with a small invariant mass. The TOF matching efficiency for electrons is then calculated as a function of electron $p_{\rm T}$, and compared to that of pions to obtain a $p_{\rm T}$-dependent electron-to-pion efficiency ratio. Finally, this $p_{\rm T}$-dependent ratio is applied as a scaling factor to the three-dimensional efficiency map for pions to obtain the electron $\varepsilon_{\rm match}$ as a function of $p_{\rm T}$, $\eta$, and $\phi$.
    
    The eID efficiency ($\varepsilon_{\rm eID}$) consists of two components: the $1/\beta$ selection efficiency and the $n\sigma_{e}$ selection efficiency. Both efficiencies are evaluated using photonic electrons from data, for which the electron sample is selected excluding the eID selection under study. The $1/\beta$ efficiency for electrons is calculated as the fraction of photonic electrons that satisfy $|1/\beta - 1|  < 0.025$ within the inclusive sample restricted to $0.5 < 1/\beta < 1.1$ to avoid hadron contamination.
    
    Similarly, the $n\sigma_{e}$ selection efficiency is obtained using the $n\sigma_{e}$ distributions of photonic electrons. Due to the excellent eID performance of TOF, the hadron contamination in the photonic electron sample based on $1/\beta$ selection is negligible. At $\sqrt{s_{\rm NN}} = 14.6, 19.6$, and 27 GeV, the distribution is well described by a Gaussian function, and the efficiency is calculated as the integral of the fitted Gaussian function within the selected range divided by the full integral. At 17.3 GeV, imperfections in the $n\sigma_e$ calibration lead to a non-Gaussian shape of the photonic-electron $n\sigma_e$ distribution, and a bin-counting method is employed instead to determine the efficiency.
    
    \section{Systematic uncertainties}
    Sources of systematic uncertainties include those arising from signal extraction and evaluations of the TPC tracking, TOF matching and eID efficiencies.
    
    For signal extraction, several variations are used: the invariant mass fitting range and histogram bin width are independently widened and narrowed, and the functional forms for the combinatorial and residual backgrounds are varied from the default to fourth- and zeroth-order polynomials, respectively. Additionally, the width of the $J/\psi$ signal shape is varied to account for uncertainties in tuning the simulation to match the momentum resolution in data. The associated uncertainty is taken as the root mean square (RMS) of all variations relative to the default value and varies across different centrality and $p_{\rm T}$ intervals. 
    
    The uncertainty in the TPC tracking efficiency is evaluated by simultaneously varying the track quality requirements in both data and simulations and finding the resulting variations in the corrected $J/\psi$ yields. To account for residual imperfections in the detector simulation, an uncertainty of 2\% per track is assigned at $\sqrt{s_{\rm NN}}$ = 14.6, 17.3, and 19.6 GeV~\cite{STAR:2025ydk}. At 27 GeV, an uncertainty of 5\% per track is assigned, as also used in the previous STAR $J/\psi$ measurement at $\sqrt{s_{\rm NN}}=54.4$ GeV~\cite{STAR:2025lxx}. The larger assigned uncertainty at 27 GeV is due to a less accurate description of the detector tracking performance in simulation compared to other energies. The largest TPC tracking-efficiency uncertainty occurs in the 2--4 GeV/$c$ interval at $\sqrt{s_{\rm NN}} = 14.6$ GeV, where the corrected yield is more sensitive to variations of the track-quality requirements due to the limited signal statistics.
    
    The uncertainty associated with the TOF matching efficiency is estimated by comparing results obtained using photonic electrons with different selection criteria for the $e^{+}e^{-}$ invariant mass. The eID-related uncertainties arise from both the $n\sigma_{e}$ and $1/\beta$ selections. The $n\sigma_{e}$ selection criteria are varied around the nominal values to evaluate the impact on the corrected $J/\psi$ yields, and the maximum deviations are taken as the uncertainties. For the $1/\beta$ selection, the uncertainty is estimated by using a different approach to mitigate hadron contamination for efficiency evaluation, {\it i.e.}, reflecting the lower-$1/\beta$ side of the electron peak about the peak center to the higher-$1/\beta$ side, where hadron contamination is larger. The difference from the nominal efficiency is taken as the systematic uncertainty.
    
    The total systematic uncertainties, obtained by adding individual sources in quadrature, range from 6.5\% to 20.6\% across different centrality and $p_{\rm {T}}$ bins, with the dominant contribution coming from the TPC tracking efficiency uncertainty. They are listed in Table \ref{tab:SysUncert}, along with individual uncertainties.
    
    \begin{table}[htbp]
    \centering
    \begin{tabular}{lr|c|c|c|c}
    \hline
    & Au+Au&14.6 GeV  &17.3 GeV &19.6 GeV &27 GeV\\
    \hline
    &signal extraction & $4.5-10.1\%$ & $4.1-8.8\%$ &$3.2-6.3\%$ & $1.8-3.8\%$ \\
    &TPC tracking & $5.3-17.1\%$ & $4.9-8.2\%$ & $4.4-7.8\%$ & $10.2-10.8\%$\\
    &TOF matching & $0.1-2.9\%$ & $0.3-1.0\%$ &$0.2-2.1\%$&$0.1-1.9\%$\\
    &$1/\beta$ selection & $1.3-2.2\%$ & $2.3-2.6\%$ & $0.8-2.0\%$ & $2.1-3.7\%$\\
    &$n\sigma_{e}$ selection & $1.7-7.3\%$ & $3.1-10.7\%$ &$1.4-6.5\%$&$3.1-8.9\%$\\
    \hline
    &Total & $8.8-20.6\%$ &$8.4-15.2\%$& $6.5-10.5\%$ &$11.5-14.4\%$\\
    \hline
    \end{tabular}
    \caption{Individual and total systematic uncertainties at different collision energies. Ranges correspond to variations across event centrality and $J/\psi$ $p_{\rm {T}}$ intervals.}
    \label{tab:SysUncert}
    \end{table}
    
    \section{Results}
\begin{figure}
    	\centering 
    	\includegraphics[width=0.6\textwidth]{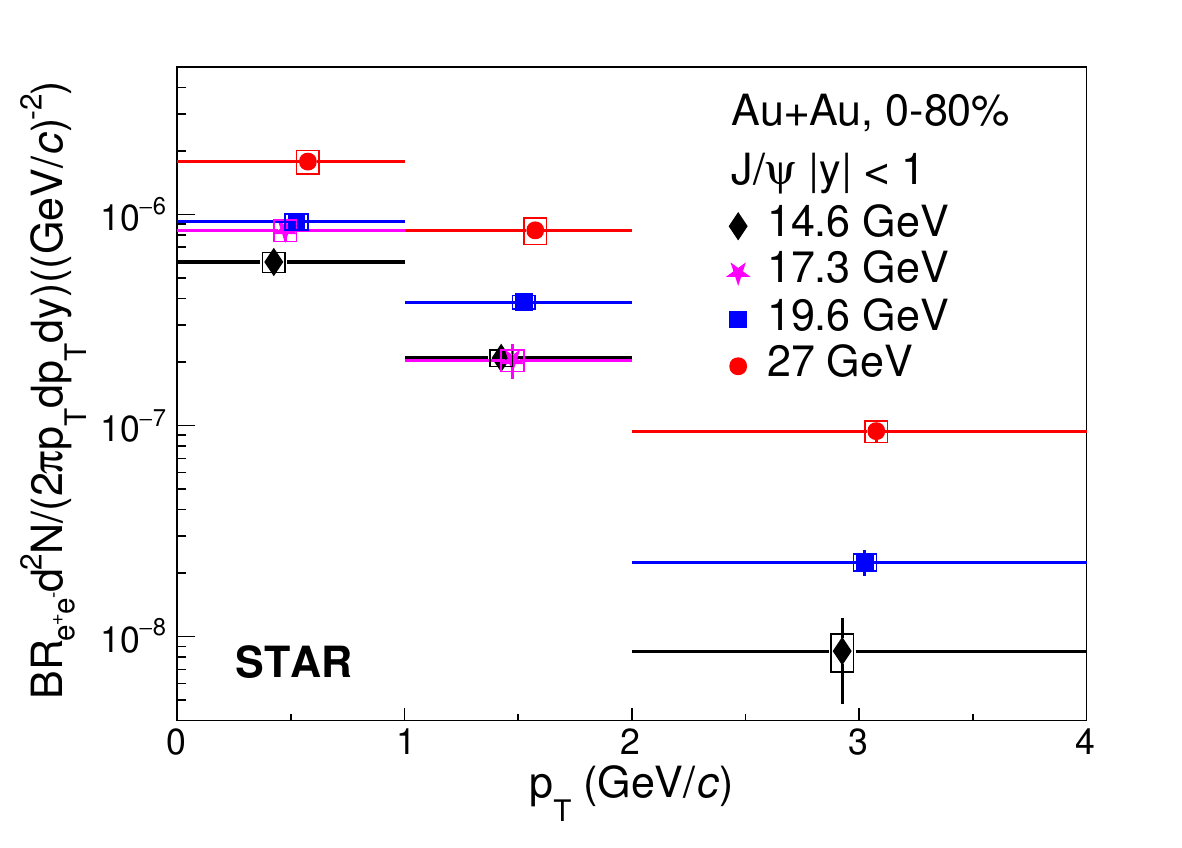}	
    	\caption{Invariant yields of inclusive $J/\psi$ as a function of $p_{\rm {T}}$ at mid-rapidity ($|y| < 1$) in Au+Au collisions at $\sqrt{s_{\rm NN}}$  = 14.6, 17.3, 19.6, and 27 GeV. The vertical bars represent statistical uncertainties, while the boxes represent systematic uncertainties. The horizontal bars indicate the $p_{\rm {T}}$ bin widths. Some data points are shifted slightly along the horizontal axis for visibility.}
    	\label{Fig:invariant_yield}%
    \end{figure}

    The invariant yields of inclusive $J/\psi$ at mid-rapidity ($|y| < 1$) in 0--80\% Au+Au collisions at $\sqrt{s_{\rm NN}} = 14.6$, 17.3, 19.6, and 27 GeV are shown in Fig.~\ref{Fig:invariant_yield} as a function of $J/\psi$ $p_{\rm {T}}$. They are calculated as:
    
    \begin{equation} BR_{J/\psi \rightarrow e^{+}e^{-}} \frac{d^2N_{J/\psi}}{2\pi p_{\rm {T}} dp_{\rm {T}} dy} = \frac{\Delta N_{J/\psi}}{2\pi p_{\rm {T}} \Delta p_{\rm {T}} \Delta y  N_{\rm {MB}}  \varepsilon_{\rm {total}}}, \label{Eq:jpsi_yield} \end{equation}
    where $\Delta N_{J/\psi}$ is the raw $J/\psi$ yield in a given $p_{\rm {T}}$ and rapidity interval ($\Delta p_{\rm {T}}$, $\Delta y$ = 2), $N_{\rm {MB}}$ is the number of minimum-bias events within 0--80\% centrality, and $\varepsilon_{\rm {total}}$ denotes the total $J/\psi$ reconstruction efficiency. For visibility, some data points in Fig.~\ref{Fig:invariant_yield} are shifted slightly along the horizontal axis, while the horizontal bars indicate the $p_{\rm T}$ bin widths. For $\sqrt{s_{\rm NN}}=17.3$ GeV, the 2--4 GeV/$c$ interval is omitted from Fig.~\ref{Fig:invariant_yield} because of the limited statistical precision of the raw $J/\psi$ yield in this interval.
    
    The nuclear modification factor $R_{{\rm AA}}$ quantifies the modification of $J/\psi$ production in heavy-ion collisions relative to that in $p$+$p$ collisions. It is defined as:
    
    \begin{equation}
    R_{\rm AA} = \frac{1}{\langle N_{\rm {coll}} \rangle / \sigma^{\rm {inel}}_{pp}}\frac{d^2N_{\rm AA}/dp_{\rm {T}} dy}{d^2\sigma_{pp}/dp_{\rm {T}} dy}
    \label{Eq:JpsiRaa},
    \end{equation}
    where $\sigma^{\rm {inel}}_{pp}$ is the $p$+$p$ inelastic cross section, ${d^2N_{\rm AA}/dp_{\rm {T}} dy}$ is the $J/\psi$ yield in A+A collisions, and ${d^2\sigma_{pp}/dp_{\rm {T}} dy}$ is the $J/\psi$ production cross section in $p$+$p$ collisions.
    
    Since no direct measurements of the $J/\psi$ cross section in $p$+$p$ collisions at $\sqrt{s} =$ 14.6, 17.3, 19.6, and 27 GeV are available, the $p$+$p$ reference values are interpolated by fitting measured inclusive $J/\psi$ cross sections over the full phase space for collision energies ranging from $\sqrt{s} = 6.8$ to $7000$ GeV~\cite{Zha:2015eca}. The fit uses the world data available in Ref.~\cite{Zha:2015eca}. The $p_{\rm {T}}$ and rapidity distributions are also parameterized based on world data, allowing for the extraction of the $p_{\rm {T}}$ differential cross section within $|y| < 1$~\cite{Zha:2015eca}. The systematic uncertainty in the $p$+$p$ reference is estimated by varying the functional forms used in the fits for the total cross section and rapidity distributions. The interpolated inclusive $J/\psi$ cross sections multiplied by the dielectron branching ratio, integrated over $p_{\rm {T}} > 0$ GeV/$c$ and $|y|<1$, in $p$+$p$ collisions at $\sqrt{s} = 14.6$, 17.3, 19.6, and 27 GeV are
    $2.09 \pm 0.40$,
    $3.36 \pm 0.56$,
    $4.70 \pm 0.55$, and
    $9.36 \pm 0.57$ nb, respectively.
    
\begin{figure}[h]
    \centering
     \includegraphics[width=1\textwidth]{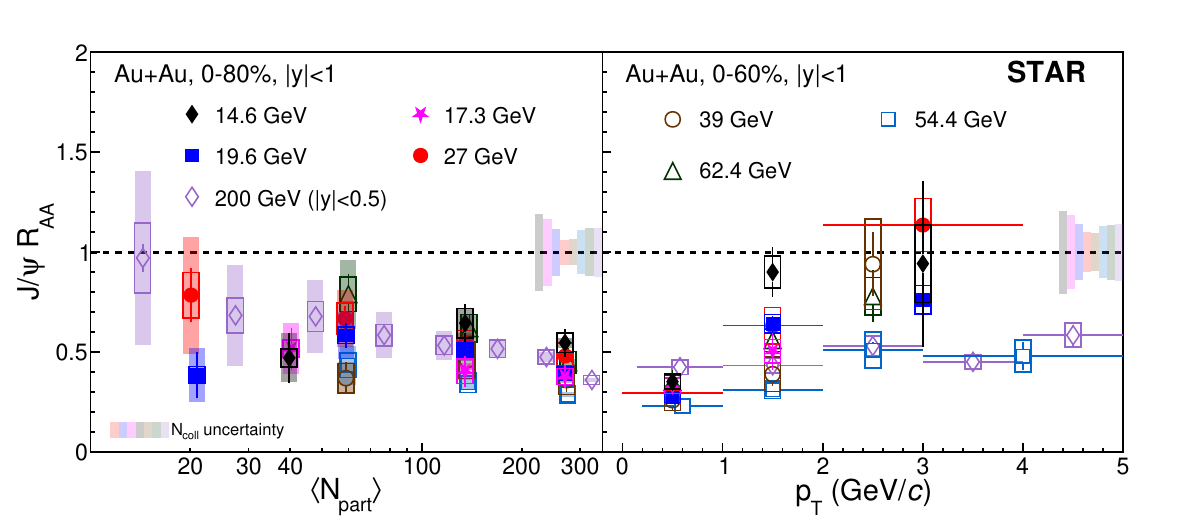}
    \caption{Inclusive $J/\psi$ $R_{\rm AA}$ at mid-rapidity ($|y|<1$) in Au+Au collisions at $\sqrt{s_{\rm NN}} =$ 14.6, 17.3, 19.6, and 27 GeV, compared with previously published results at 39, 54.4, 62.4, and 200 GeV~\cite{STAR:2016utm,STAR:2025lxx,STAR:2019fge}. The left panel shows the centrality dependence integrated over $p_{\rm T}$, while the right panel shows the $p_{\rm T}$ dependence integrated over 0--80\% or 0--60\%. Vertical bars and open boxes represent statistical and systematic uncertainties, respectively. In the left panel, the filled boxes around data points denote the uncertainty in $\langle N_{\rm coll} \rangle$, while the semi-transparent boxes around unity indicate the global uncertainties from the $p$+$p$ reference. In the right panel, the semi-transparent boxes around unity indicate the combined global uncertainties from the $p$+$p$ reference and $\langle N_{\rm coll} \rangle$. The global uncertainties are arranged from left to right according to increasing collision energy.}
    
    \label{Fig:Raa_BESII}       
    \end{figure}

    
    The centrality dependence of the inclusive $J/\psi$ $R_{\rm AA}$ is shown in the left panel of Fig.~\ref{Fig:Raa_BESII}, where data are presented as a function of $\langle N_{\rm part} \rangle$ for Au+Au collisions at $\sqrt{s_{\rm NN}} =$ 14.6, 17.3, 19.6, and 27 GeV. The measurements are performed at mid-rapidity ($|y|<1$) and compared with previously published results at 39, 54.4, and 62.4 GeV~\cite{STAR:2016utm,STAR:2025lxx} as well as 200 GeV ($|y|<0.5$)~\cite{STAR:2019fge}. The difference in rapidity coverage for the 200 GeV measurement should have a small impact on the comparison, since the medium effect at this energy is not expected to change significantly within $|y|<1$. In the new measurements, $R_{{\rm AA}}$ values are around 0.5 in 0--20\% central collisions, suggesting a suppression of inclusive $J/\psi$ production at collision energies between 14.6 and 27 GeV. A hint of a decreasing trend for $R_{{\rm AA}}$ from peripheral to central collisions is observed, qualitatively consistent with the pattern seen at 39--200 GeV. 
    
    The $p_{\rm T}$ dependence of $J/\psi$ $R_{{\rm AA}}$, integrated over centrality, is shown in the right panel of Fig.~\ref{Fig:Raa_BESII} for Au+Au collisions at $\sqrt{s_{\rm NN}} =$ 14.6--27 GeV, providing complementary information on the interplay between different effects. The measured $R_{{\rm AA}}$ values show a tendency to increase with increasing $J/\psi$ $p_{\rm {T}}$. Such a behavior is consistent with an important role of CNM effects at these collision energies, for which the yield suppression is expected to be more pronounced at low $p_{\rm T}$ and to decrease toward higher $p_{\rm T}$~\cite{Nagle:2010ix,Lourenco:2008sk,Andronic:2025jbp}. At $p_{\rm {T}} >$ 2 GeV/$c$, the measurements are consistent with unity within uncertainties. Similar measurements in 39, 54.4, 62.4, and 200 GeV Au+Au collisions are also shown in the right panel of Fig.~\ref{Fig:Raa_BESII} for comparison~\cite{STAR:2016utm,STAR:2025lxx,STAR:2019fge}. The $p_{\rm T}$ dependence of $R_{{\rm AA}}$ seems to be different between 200 GeV and 14.6--27 GeV. This difference could arise from a stronger regeneration contribution at 200 GeV in the low $p_{\rm {T}}$ region, which compensates for the dissociation and CNM effects, and a stronger dissociation effect at higher $p_{\rm {T}}$, compared to those at low energies. The strong suppression above 2 GeV/$c$ is also present at 54.4 GeV.
    
\begin{figure}
    	\centering 
    	\includegraphics[width=0.9\textwidth]{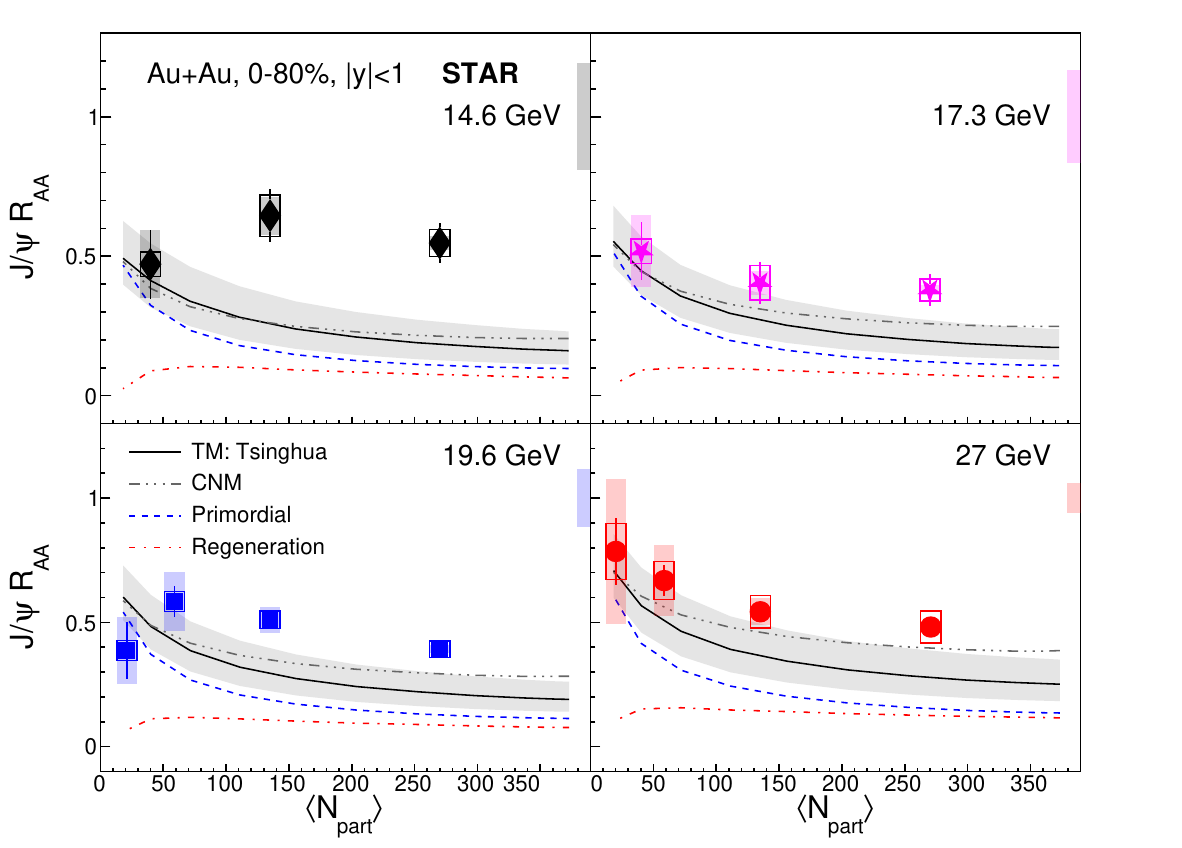}	
    \caption{$J/\psi$ $R_{\rm AA}$ as a function of $\langle N_{\rm part} \rangle$ at mid-rapidity ($|y|<1$) in Au+Au collisions at $\sqrt{s_{\rm NN}} = 14.6$, 17.3, 19.6, and 27 GeV. Data points represent STAR measurements. Vertical bars on data points show statistical uncertainties, while the surrounding boxes and shaded bands represent systematic uncertainties arising from the Au+Au analysis and $\langle N_{\rm coll} \rangle$, respectively. Vertical bands at unity show the uncertainties in the $p$+$p$ reference. Curves show calculations from the Tsinghua transport model~\cite{Helenius:2012ny,Lourenco:2008sk,Zhao:2020jqu,Zhao:2022ggw,Zha:2015eca}, with different components illustrated by different curve styles. The shaded bands around the model curves represent model uncertainties.}
    	\label{Fig:RaaVsNpart_with_Theory}%
    \end{figure}

\begin{figure}
    	\centering 
    	\includegraphics[width=0.9\textwidth]{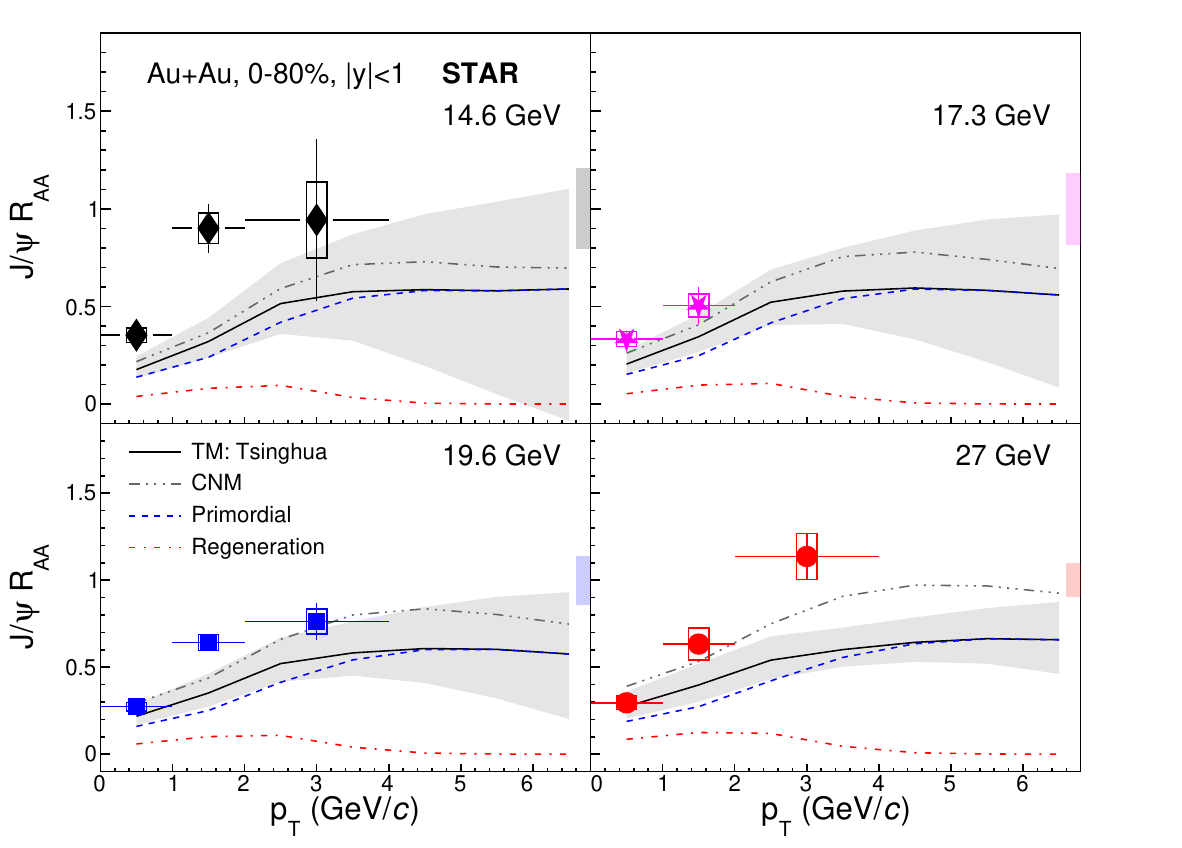}	
    \caption{$J/\psi$ $R_{\rm AA}$ as a function of $p_{\rm {T}}$ at mid-rapidity ($|y|<1$) in Au+Au collisions at $\sqrt{s_{\rm NN}} = 14.6$, 17.3, 19.6, and 27 GeV. Data points represent STAR measurements. Vertical bars on data points show statistical uncertainties, while the surrounding boxes represent systematic uncertainties arising from the Au+Au analysis. Vertical bands at unity show the combined global uncertainties from the $p$+$p$ reference and $\langle N_{\rm coll} \rangle$. Curves show calculations from the Tsinghua transport model~\cite{Helenius:2012ny,Lourenco:2008sk,Zhao:2020jqu,Zhao:2022ggw,Zha:2015eca}, with different components illustrated by different curve styles. The shaded bands around the model curves represent model uncertainties.}
    	\label{Fig:RaaVspt_with_Theory}%
    \end{figure}

    Detailed comparisons of the $J/\psi$ $R_{\mathrm{AA}}$ as a function of $\langle N_{\rm part} \rangle$ and $p_{\rm T}$ with transport model (TM) calculations from the Tsinghua group~\cite{Helenius:2012ny,Lourenco:2008sk,Zhao:2020jqu,Zhao:2022ggw,Zha:2015eca} are presented in Figs.~\ref{Fig:RaaVsNpart_with_Theory} and~\ref{Fig:RaaVspt_with_Theory} for Au+Au collisions at $\sqrt{s_{\rm NN}} =$ 14.6--27 GeV. These calculations include primordially produced $J/\psi$ (including those produced directly and from decays of excited charmonium states) and regenerated $J/\psi$, while the non-prompt component from $B$-hadron decays is not included. At the top RHIC energy, the non-prompt contribution to the inclusive $J/\psi$ yield is expected to be at the few-percent level~\cite{Andronic:2015wma,Cacciari:2005rk}, and is expected to be further reduced at BES-II energies. Different components in the model are shown separately: i) CNM effects arising from nPDF effects based on the EPS09 parameterization~\cite{Helenius:2012ny}, nuclear absorption~\cite{Lourenco:2008sk} and the Cronin effect~\cite{Cronin:1974zm}, ii) primordial $J/\psi$ that undergo CNM effects and in-medium dissociation, and iii) regenerated $J/\psi$. The uncertainties of the Tsinghua calculation, shown as shaded bands, are obtained from three sources: the charm quark cross section, the nPDFs, and nuclear absorption. Overall, the transport model calculations are broadly compatible with the measured centrality and $p_{\rm T}$ dependences within uncertainties. For the centrality dependence, a noticeable deviation is found at $\sqrt{s_{\rm NN}}=14.6$~GeV, where the measured values at larger $\langle N_{\rm part} \rangle$ are above the calculation. For the $p_{\rm T}$ dependence, the calculations describe the qualitative trend but tend to lie below the measured values in some higher-$p_{\rm T}$ intervals. 
    
\begin{figure}[h]
    
    \centering
    \includegraphics[width=1\textwidth]{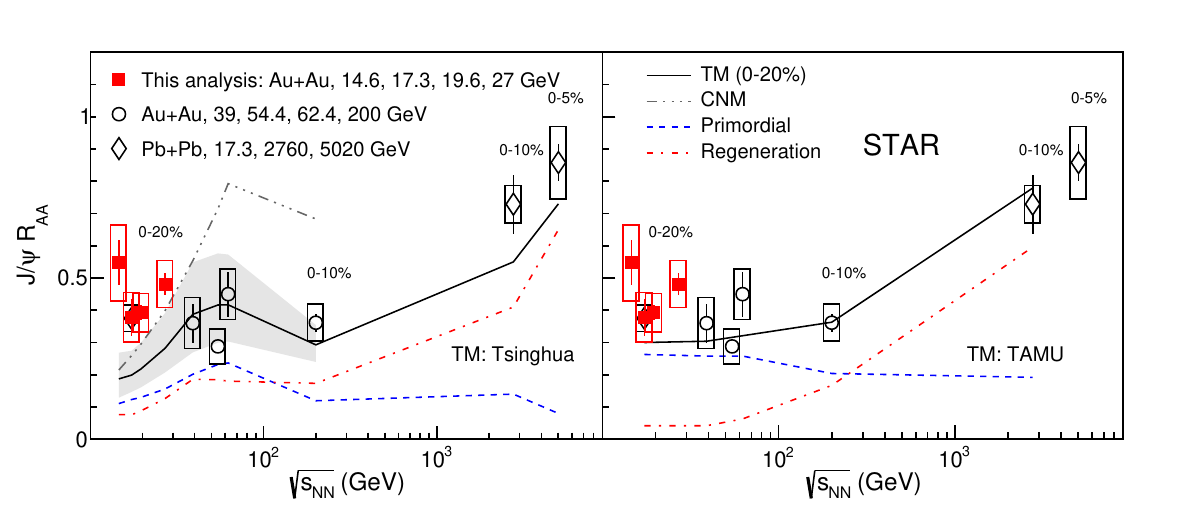}
    \caption{$J/\psi$ $R_{{\rm AA}}$ as a function of collision energy in central collisions~\cite{NA50:2000brc,Kluberg:2005yh,ALICE:2013osk,STAR:2016utm,STAR:2025lxx,STAR:2019fge,Bai:2020svs}, compared with two transport model (TM) calculations from the Tsinghua group (left)~\cite{Helenius:2012ny,Lourenco:2008sk,Zhao:2020jqu, Zhao:2022ggw, Zha:2015eca} and the TAMU group (right)~\cite{Zhao:2010nk}. Vertical bars represent statistical uncertainties, while the open boxes around data points represent systematic uncertainties, including contributions from the Au+Au analysis, the $p$+$p$ reference, and $\langle N_{\rm coll} \rangle$. For the Tsinghua calculation, the shaded band around the model curve represents the uncertainty.}
    \label{Fig:RaaVsEnergy}       
    \end{figure}

    The collision energy dependence of inclusive $J/\psi$ $R_{{\rm AA}}$ in central collisions is examined in Fig.~\ref{Fig:RaaVsEnergy}, combining the new measurements at 14.6--27 GeV with previously published results~\cite{NA50:2000brc,Kluberg:2005yh,ALICE:2013osk,STAR:2016utm,STAR:2019fge,Bai:2020svs}. Within uncertainties, the data do not show a clear energy dependence below 200 GeV, likely arising from an interplay of different effects. At 200 GeV, the suppression is generally attributed to dissociation processes, with a non-negligible contribution from regeneration~\cite{STAR:2019fge,Zhao:2010nk}. In contrast, at 27 GeV and below, the medium temperature and charm quark cross section are significantly reduced compared to those at 200 GeV. This is expected to result in weaker dissociation and regeneration effects~\cite{Andronic:2025jbp,Zhao:2020jqu,Zhao:2022ggw}. On the other hand, CNM effects are expected to increase with decreasing collision energy, mainly due to enhanced nuclear absorption from colliding nuclei spending a longer time passing through each other. Transport model calculations from the Tsinghua group~\cite{Helenius:2012ny,Lourenco:2008sk,Zhao:2020jqu, Zhao:2022ggw, Zha:2015eca} and the TAMU group~\cite{Zhao:2010nk} are also shown in Fig.~\ref{Fig:RaaVsEnergy} for comparison. In both cases, model predictions for primordial and regenerated $J/\psi$ production are shown separately, and the non-prompt contribution is not included. Both transport models qualitatively describe the overall trend of the data, with the predicted $R_{{\rm AA}}$ values being systematically lower than the measurements at 27 GeV and below. A peak structure around $\sqrt{s_{\rm NN}} \sim 50$ GeV appears in the Tsinghua model, originating from the interplay of the energy dependence of nPDF effects and nuclear absorption. 
    
    
    In addition to $R_{{\rm AA}}$, one can also use $R_{\rm CP}$, the relative modification of $J/\psi$ yields in central collisions compared to those in peripheral collisions, to study how the QGP affects $J/\psi$ production. It is defined as:
    \begin{equation} R_{\rm CP} = \frac{\left( \frac{dN_{\rm AA}/dy}{\langle N_{\rm {coll}} \rangle} \right)_{\rm {central}}}{\left( \frac{dN_{\rm AA}/dy}{\langle N_{\rm {coll}} \rangle} \right)_{\rm {peripheral}}}, \end{equation} 
    where $dN/dy$ represents the $J/\psi$ yield in a given centrality class, and $\langle N_{\rm {coll}} \rangle$ is the average number of binary collisions for that class. An advantage of using $R_{\rm CP}$ over $R_{\rm{AA}}$ is that it does not require a $p$+$p$ reference, thereby avoiding the uncertainties associated with the interpolation procedure. Furthermore, correlated systematic uncertainties in the $J/\psi$ yield measurements between central and peripheral collisions cancel out in $R_{\rm CP}$. Specifically, the systematic uncertainties of $R_{\rm CP}$ are calculated as follows: (1) signal extraction: treated as uncorrelated and combined in quadrature; (2) tracking efficiency: considered fully correlated and therefore largely canceling out in $R_{\rm CP}$; (3) eID efficiency: fully correlated but only partially canceling out due to detector performance variations across centrality; and (4) $\langle N_{\rm {coll}} \rangle$ uncertainty: fully correlated and partially canceling out. The total systematic uncertainty is then obtained by adding the individual contributions in quadrature.
    
\begin{figure}[h]
    
    \centering
    \includegraphics[width=1\textwidth]{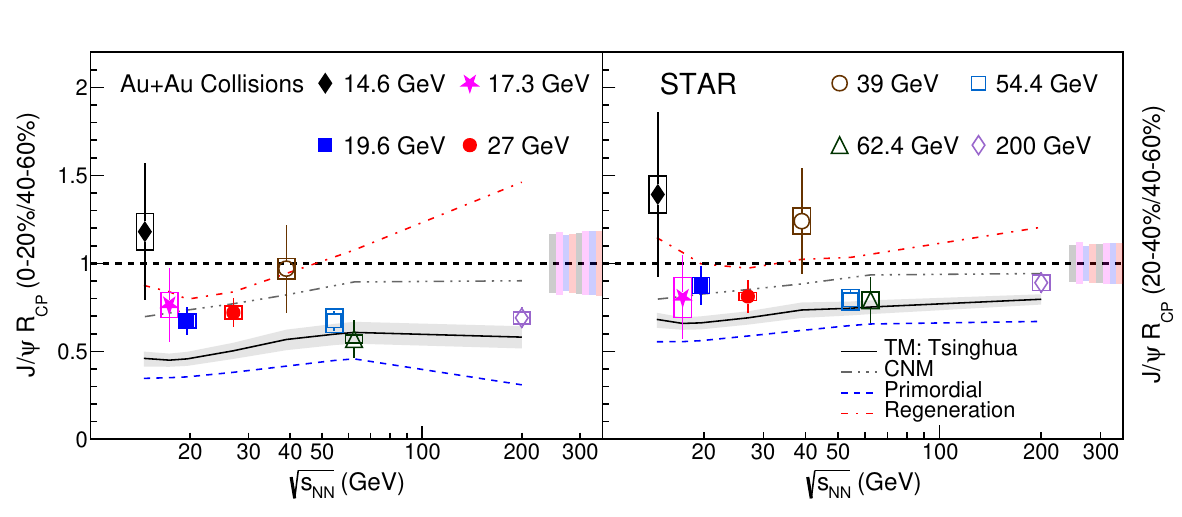}
    \caption{$J/\psi$ $R_{\rm CP}$ as a function of collision energy for the 0--20\% (left) and 20--40\% (right) centralities, with 40--60\% peripheral events taken as the reference. Vertical bars denote statistical uncertainties, and open boxes represent systematic uncertainties. Colored bands around unity indicate the uncertainties in $\langle N_{\rm {coll}} \rangle$ for 14.6 GeV to 200 GeV from left to right. Transport model calculations by the Tsinghua group~\cite{Helenius:2012ny,Lourenco:2008sk,Zhao:2020jqu, Zhao:2022ggw, Zha:2015eca} are shown for comparison, with uncertainties shown as shaded bands.}
    \label{Fig:RcpVsEnergy}
    \end{figure}

    Figure~\ref{Fig:RcpVsEnergy} shows $R_{\rm CP}$ as a function of collision energy in Au+Au collisions for the 0--20\% and 20--40\% centrality classes, including the new BES-II measurements and previously published results~\cite{STAR:2016utm,STAR:2025lxx}. The 40--60\% centrality class is chosen as the reference to be consistent with previous STAR measurements and to balance statistical precision with the magnitude of remaining medium effects. To quantify the energy dependence, the $R_{\rm CP}$ results are fitted with a linear function of $\sqrt{s_{\rm NN}}$. The extracted slopes are $(-7.3 \pm 4.9)\times 10^{-5}$~GeV$^{-1}$ and $(2.9 \pm 5.5)\times 10^{-4}$~GeV$^{-1}$ for the 0--20\% and 20--40\% centrality classes, respectively, with $\chi^{2}/{\rm ndf}=4.1/6$ and $3.7/6$. The fitted slopes do not provide significant evidence for a collision-energy dependence over the studied energy range. A constant fit is further applied to quantify the suppression. The fitted values are $0.64 \pm 0.04~(\mathrm{fit}) \pm 0.11~(\mathrm{global})$ and $0.85 \pm 0.04~(\mathrm{fit}) \pm 0.10~(\mathrm{global})$ for the 0--20\% and 20--40\% centrality classes, respectively, suggesting stronger relative suppression in more central collisions. Transport model calculations by the Tsinghua group~\cite{Helenius:2012ny,Lourenco:2008sk,Zhao:2020jqu, Zhao:2022ggw, Zha:2015eca} are shown for comparison, with the uncertainty shown as shaded bands. While model calculations tend to lie below measured $R_{\rm CP}$ central values at $\sqrt{s_{\rm NN}} \leq 39$ GeV, they are still consistent with data within experimental and model uncertainties.

     \section{Summary}
    In summary, measurements of inclusive $J/\psi$ production via the dielectron channel at mid-rapidity ($|y|<1$) in Au+Au collisions at $\sqrt{s_{\rm NN}} = 14.6$, 17.3, 19.6, and 27 GeV are presented by the STAR experiment at RHIC. These results explore a collision energy regime in which previous measurements were limited. A suppression of $J/\psi$ yields is observed in central collisions at all measured energies, with no significant energy dependence up to 200 GeV within uncertainties. Furthermore, $J/\psi$ $R_{\rm{AA}}$ shows a tendency to increase toward higher $J/\psi$ $p_{\rm {T}}$ in the new measurements. Transport model calculations, which account for CNM, dissociation, and regeneration effects, qualitatively describe the overall trend of the energy dependence. At $\sqrt{s_{\rm NN}} \leq 27$ GeV, however, model calculations appear to predict lower $R_{\rm AA}$ values than measured in central collisions. These results provide new constraints on the interplay of various medium effects influencing $J/\psi$ production and advance the extraction of QGP properties from $J/\psi$ measurements.

    \acknowledgments
    We thank the RHIC Operations Group and SCDF at BNL, the NERSC Center at LBNL, and the Open Science Grid consortium for providing resources and support. This work was supported in part by the Office of Nuclear Physics within the U.S. DOE Office of Science, the U.S. National Science Foundation, National Natural Science Foundation of China, Chinese Academy of Sciences, the Ministry of Science and Technology of China and the Chinese Ministry of Education, NSTC Taipei, the National Research Foundation of Korea, Czech Science Foundation and Ministry of Education, Youth and Sports of the Czech Republic, Hungarian National Research, Development and Innovation Office, New National Excellency Programme of the Hungarian Ministry of Human Capacities, Department of Atomic Energy and Department of Science and Technology of the Government of India, the National Science Centre and WUT ID-UB of Poland, German Bundesministerium f\"ur Bildung, Wissenschaft, Forschung and Technologie (BMBF), Helmholtz Association, Ministry of Education, Culture, Sports, Science, and Technology (MEXT), Japan Society for the Promotion of Science (JSPS), and Agencia Nacional de Investigacion y Desarrollo de Chile (ANID), Chile.
    

    
    \bibliographystyle{JHEP}
    \bibliography{biblio.bib}
    
    
    \end{document}